\documentclass[reprint,aps,prb,twocolumn,superscriptaddress]{revtex4-1}
\usepackage[utf8]{inputenc}
\usepackage{graphicx}
\usepackage{dcolumn}
\usepackage{xcolor}
\usepackage{bm}
\usepackage{amsmath}
\usepackage[english]{babel}
\usepackage[utf8]{inputenc}
\usepackage{color}
\begin{document}

\title{A quantum dot crossbar with sublinear scaling \\of interconnects at cryogenic temperature}
\author{P. L. Bavdaz}
\author{H. G. J. Eenink}
\author{J. van Staveren}
\author{M. Lodari}
\affiliation{QuTech and Kavli Institute of Nanoscience, Delft University of Technology, PO Box 5046, 2600 GA Delft, The Netherlands}
\author{C.~G.~Almudever}
\affiliation{Computer Engineering Department, Technical University of Valencia, Camino de Vera, s/n, 46022 València, Spain}
\author{J. S. Clarke}
\affiliation{Intel Corporation, Technology and Manufacturing Group, Hillsboro, Oregon, 97124, USA}
\author{F.~Sebastiano}
\author{M. Veldhorst}
\author{G. Scappucci}
\affiliation{QuTech and Kavli Institute of Nanoscience, Delft University of Technology, PO Box 5046, 2600 GA Delft, The Netherlands}
\date{\today}

\begin{abstract}
We demonstrate a 36$\times$36 gate electrode crossbar that supports 648 narrow-channel field effect transistors (FET) for gate-defined quantum dots, with a quadratic increase in quantum dot count upon a linear increase in control lines. The crossbar is fabricated on an industrial $^{28}\text{Si}$-MOS stack and shows 100\% FET yield at cryogenic temperature. We observe a decreasing threshold voltage for wider channel devices and obtain a normal distribution of pinch-off voltages for nominally identical tunnel barriers probed over 1296 gate crossings. Macroscopically across the crossbar, we measure an average pinch-off of 1.17~V with a standard deviation of 46.8~mV, while local differences within each unit cell indicate a standard deviation of 23.1~mV. These disorder potential landscape variations translate to 1.2 and 0.6 times the measured quantum dot charging energy, respectively. Such metrics provide means for material and device optimization and serve as guidelines in the design of large-scale architectures for fault-tolerant semiconductor-based quantum computing.

\end{abstract}

\maketitle

Semiconductor spin-qubits in gate-defined quantum dots are promising building blocks for quantum computers\cite{Vandersypen2017}. Spin qubit can exhibit long quantum coherence times\cite{Veldhorst2014}, can be operated with high-fidelity single and two-qubit logic\cite{Yoneda2018, Yang2019a, Lawrie2021, Xue2022, Noiri2022}, can be operated at comparatively high temperature\cite{Yang2020,Petit2020}, and can be fabricated using semiconductor manufacturing\cite{Maurand2016, Zwerver2021}. Building upon this, recent devices have been scaled to contain up to 9 dots in a linear array\cite{Mills2019a} and a universal four-qubit quantum processor positioned in a 2$\times$2 array\cite{Hendrickx2021}.

However, a practical spin-based quantum computer will require orders of magnitude more qubits. Qubits operating at cryogenic temperature will have to interface with room temperature control. Brute-force approaches where the number of control lines to room temperature scale with the number of qubits will become unsustainable\cite{Franke2019}. Instead, architectures have been proposed\cite{Veldhorst2017,Li2018,Taylor2005} that allow for sublinear interconnect scaling, relying on challenging levels of device uniformity and integration of cryogenic electronics\cite{Xue2021}. Simultaneously keeping all quantum dots within the desired charge state\cite{Li2018} requires either immaculate material and fabrication or unique potentials applied to each dot. These requirements can be alleviated by omitting inter-unit cell coupling from the architecture, such that shared control can be implemented to achieve sublinear interconnect scaling with current technology. The resulting unit cells are simplified and well suited for investigating the reproducibility requirements needed to operate the previously mentioned architectures. Furthermore, the sublinear scaling of interconnects enables a high-throughput fabrication-measurement cycle of quantum devices that can be used to quantify and improve device uniformity.

Crossbars have been successfully applied for high throughput measurements of quantum devices\cite{Al-Taie2013} by integrating into the device design an on-chip multiplexer specifically designed for characterization. Instead, off-chip cryogenic CMOS multiplexer platforms have been developed to increase measurement throughput of quantum devices, agnostic with respect to device design\cite{Pauka2021,PaqueletWuetz2020}. However, the lines between multiplexer and device scale linearly and remain a potential I/O bottleneck. Here we demonstrate the scalable addressability of a quantum dot crossbar architecture operated with an off-chip cryogenic multiplexer. We gather statistical data on narrow channel field effect transistors (FET) with tunable tunnel barriers, which act as the unit cells of a 2D crossbar. By introducing interleaved ohmic contacts, we address each unit cell individually with no shared current paths through the 2D electron gas (2DEG), allowing for a direct comparison between unit cells. By making use of cryo-CMOS electronics to further reduce the interconnects, we measure up to 648 FETs in a single cooldown. All together, this design establishes a powerful yet simple tool for targeting the reproducibility challenge that is crucial for realizing spin-based quantum computers.

\begin{figure}%
    \includegraphics[width=89mm]{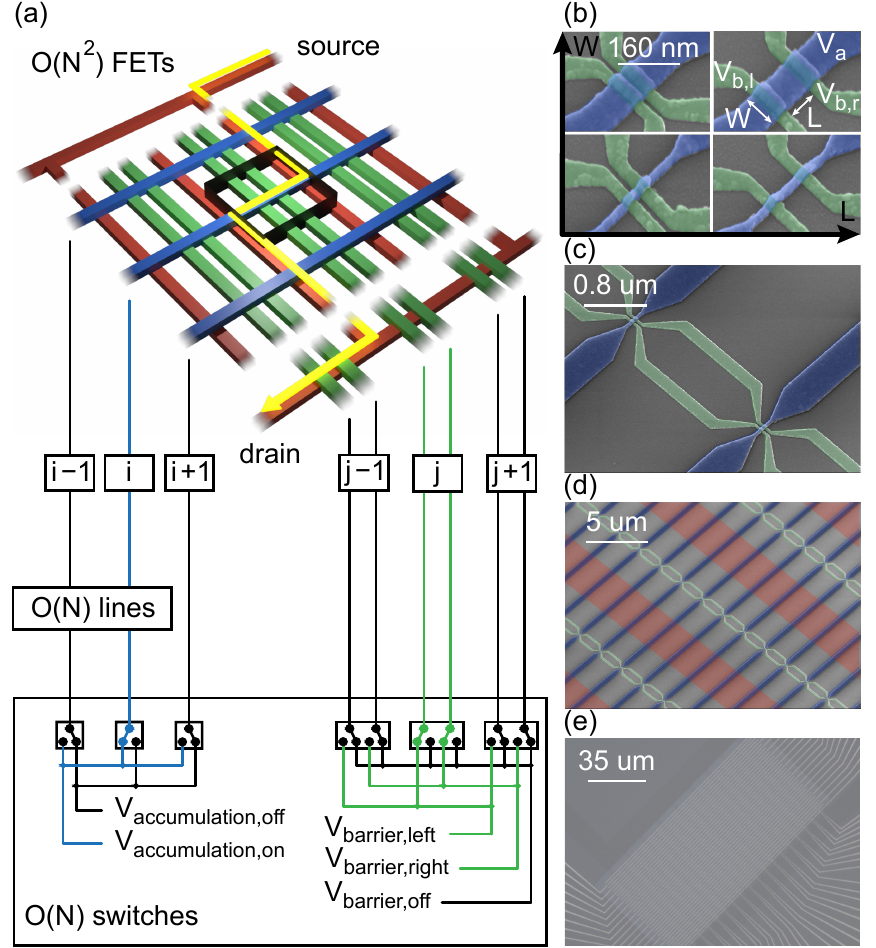}%
    \caption{(a) Schematic representation of the quantum dot crossbar connected to cryo-CMOS multiplexing control circuitry. A sub-group of 3$\times$3 unit cells hosting multi-gate field effect transistors (FET) is shown (top panel), with interleaved implant regions for source/drain ohmic contacts ($n^{++}$, red), rows of accumulation gates (AG, blue) to define a conductive channel, and columns of barrier gates (BG, green) to tune transport into the single electron regime. Indices $i$ and $j$ correspond to AG rows and BG pair columns across the crossbar array. Cryo-CMOS multiplexers on a printed circuit board (bottom panel) are used to route voltage biases and select the desired AG row and BG column, thereby creating a unique current path (yellow arrow) through a single unit cell (indicated by the black border) comprising one AG and two BGs and identified with index pair ($i,j$). With a linear increase in wires and switches, the number of measurable FETs scales quadratically. (b) False-colored scanning electron microscope (SEM) micrographs of four FETs, with the maximal variations of the AG width $W$ and the distance between BGs $L$. These FETs are located at the four corners of a $36\times36$ electrode crossbar fabricated on a $^{28}\text{Si}/\text{SiO}_2$ stack. The AG and BGs are represented in blue and green, respectively. (c) Two FETs with shared barrier gates, featuring gate widening to aid lift-off. (d) Zoomed-out SEM of multiple unit cells, including implanted ohmics represented in red. (e) Optical microscope image of the crossbar featuring 648 multi-gate FETs and 1296 crossings between AG and BG with contact fan-out lines.}
\label{fig:setup}
\end{figure}

Our experimental setup (Fig.~\ref{fig:setup}a, schematics) consists of a crossbar of multi-gate field effect transistors and a cryo-CMOS multiplexing circuit. At the heart of the crossbar is the unit cell which contains a single FET (indicated by the black border). The FET comprises an accumulation gate (AG, blue), two barrier gates (BG, green) perpendicular to the AG, and ohmic contacts (red) on either side of the AG. Electrical transport through the FET is achieved by accumulating a 2D electron gas (2DEG) channel defined under the AG and between source and drain ohmics using voltages applied to the AG. With the BGs we form tunnel barriers, capable of tuning transport into the single electron regime.

The FETs discussed in this work are fabricated on an isotopically enriched $^{28}\text{Si}/\text{SiO}_2$ stack deposited on 300 mm Si wafers in an industrial CMOS fab\cite{Sabbagh2019}, featuring a 10 nm thick thermal $\text{SiO}_2$ oxide. The device is fabricated in an academic clean room environment via electron beam lithography and electron beam evaporation. First $P^{+}$ is implanted to produce the ohmics. Next, the 20 nm thick $\text{Ti}/\text{Pd}$ BGs are fabricated. Then, a 14~nm thick $\text{Al}_2\text{O}_3$ layer is deposited via atomic layer deposition which isolates the barrier gates from the 40 nm thick $\text{Ti}/\text{Pd}$ AG. The same material stack has been used to fabricate individual quantum dot\cite{Petit2018} and qubit devices\cite{Petit2020}, and understanding the uniformity is key toward scaling beyond these devices. While our demonstration uses Si-MOS, the crossbar design can be adapted to other accumulation-mode material stacks and is thereby compatible with the leading platforms for semiconductor quantum technology. The gate layouts of four FETs with differing AG width $W$ and distance between the BGs $L$ are depicted in the first scanning electron microscope (SEM) micrographs of Figure~\ref{fig:setup}b. 

The unit cells in the crossbar, identified by index pairs ($i,j$), share gates and ohmic contacts with neighboring unit cells. Figures \ref{fig:setup}c-e show increasingly zoomed out micrographs of the crossbar with false coloring that highlight the shared gates and ohmics. Each row of unit cells ($i,*$) shares the same AG, while each column ($*,j$) shares its two BGs. The two ohmic contacts are instead shared by all unit cells and are positioned at the top and bottom of the crossbar. To allow for independent operation of each unit cell, the ohmic contacts are extended between each vertical column of unit cells, alternating between the top and bottom ohmic contacts. Figure~\ref{fig:setup}e shows an optical microscope image of the entire fabricated grid which features a total of 36 AGs and 36 BGs, thus having 1296 gate crossings and 648 FET unit cells. The intended current flow through the shared ohmics and a single unit cell is indicated by the yellow arrow in Fig.~\ref{fig:setup}a. The crossbar is sparse to prevent electrical shorts between neighboring source-drain implant extensions due to lateral implant diffusion, with a minimum distance between implanted regions of 7.5 $\mu$m.

Thanks to this crossbar approach, different unit cell designs can be explored and evaluated by introducing incremental design differences across the grid. Here we designed each unit cell with a unique combination of AG width $W = 30 + 3 \cdot i$~nm and distance between the BGs $L = 30 + 6 \cdot j$ nm, with $i = 0, ..., 35$ and $j = 0, ..., 17$. As a result, both $W$ and $L$ range from 30 to 130~nm, which are typical dimensions for quantum dots and single electron transistors (SET) in Si-MOS, Si/SiGe and Ge/SiGe\cite{Petit2020,Lawrie2020}. Figure~\ref{fig:setup}b shows FETs at the four corners of the grid, with the extreme variations of $W$ and $L$.

Control circuitry, schematically depicted in Fig.~\ref{fig:setup}a, is essential to select the unit cell under test and multiply the lines available at cryogenic temperatures. The crossbar and the cryo-CMOS control circuitry are hosted on separate printed circuit boards, connected through a flex-cable for modularity and fast sample exchanges. Both printed circuit boards are cooled to a base temperature of $1.7$ K in a variable temperature insert cryostat. The circuit contains classical CMOS single-pole-double-throw switches and its design is based on previous work\cite{PaqueletWuetz2020}, where each input terminal is connected to the sample and the output terminals are used to apply or measure voltages with room temperature equipment.

A key difference compared to previous work~\cite{PaqueletWuetz2020} is the addition of independent groups of switches. The shift register outputs are separated into two distinct groups, intended for independent control of two categories of gate. In our case we separate horizontal AG and vertical BG. As each unit cell contains one AG and two BGs, each horizontal and vertical output bit should at least control 1 and 2 switches, respectively. The number of unit cells in a crossbar with $N$ switches is maximized when the ratio of switches for accumulation gates to barrier gates is 1:1, resulting in $N^2/8$ unit cells. Our implementation of the multiplexer PCB uses 72 switches to address the 36 AGs and 18 BG pairs, resulting in 648 unit cells. To allow for more advanced unit cells in the future, each horizontal and vertical output bit controls more than 1 and 2 switches, respectively, at the cost of a constant number of additional lines to room temperature. By grounding the additional gates selected by the same bits that do not contribute to the unit cell under test, undesired current paths through other unit cells can be avoided and the setup can be operated as depicted in Fig.~\ref{fig:setup}a. Due to the proposed design with shared gates and ohmics, all selected FETs in each row (column) have identical accumulation (barrier) gate voltages and could be measured as a parallel circuit. However, for the purpose of comparing FETs and gathering statistics, single FETs are selected and measured individually in the following analysis. As switching is exclusively performed between measurements, the switching frequency-dependent power consumption of the multiplexer\cite{PaqueletWuetz2020} is never a concern.

\begin{figure}%
    \includegraphics[width=89mm]{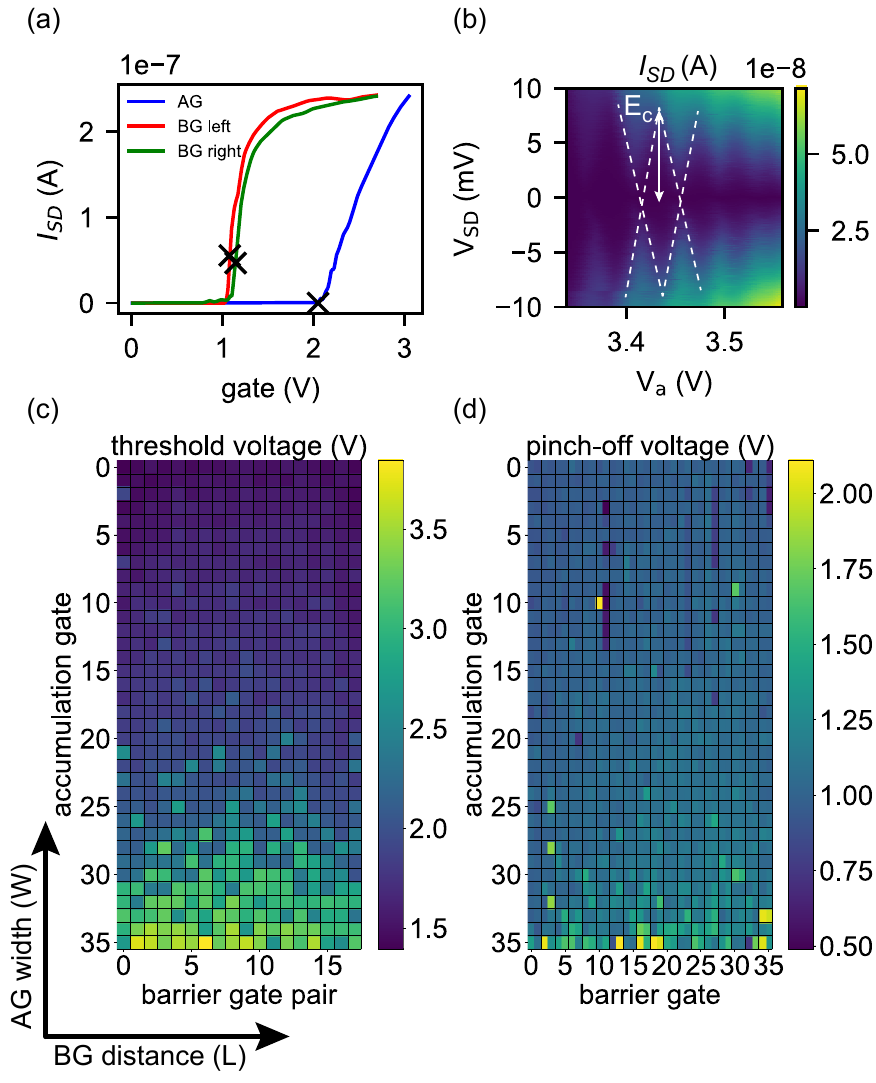}%
    \caption{(a) Source-drain current $I_{SD}$ through a selected field effect transistor measured at $T = 1.7$~K as a function of accumulation gate voltage $V_a$ with fixed barrier voltages $V_b = 2.7$~V (blue line). The turn on threshold voltage $V_{to}$ is indicated with a black cross. The same transistor is measured as a function of $V_{b,left}$ and $V_{b,right}$ at $1$~V above $V_{to}$ (red and green lines), with black crosses indicating pinch-off voltages $V_{po}$. (b) $I_{SD}$ as a function of source-drain bias $V_{SD}$ and $V_a$ with fixed $V_b$ measured at $T = 1.7$~K. In the single electron transport regime, Coulomb diamonds are observed (dashed white lines) from which the charging energy $E_c = 8$~meV and the lever arm $\alpha = 0.2$ eV/V can be extracted. (c-d) Color-scale maps of all measured $V_{to}$ (c) and $V_{po}$ (d) in a grid of varying quantum dot widths $W$ and lengths $L$ as a function of location within the grid, where each box represents an FET unit cell in the array. Each box in (d) is split in two to indicate both $V_{po,left}$ and $V_{po,right}$ values. }
\label{fig:measurement}
\end{figure}

To select a unit cell and allow current to flow from drain to source ohmics through the 2DEG, positive voltages are applied to one AG and to one BG pair as the fabricated crossbar is designed to work in accumulation mode. Current can only flow through the unit cell at the intersection of these gates since all other unselected gates associated with other cells are grounded. The grounded BGs interrupt the 2DEG thanks to electric field screening, because these gates are situated under the AG. Therefore, a unit cell can be turned-on and pinched-off using the AG and BGs, respectively, without interference from current flowing through the 2DEG in any other unit cell. 

Figure~\ref{fig:measurement}a shows an example of such measurements of an individual FET in the crossbar and how the associated key metrics at $T = 1.7$~K are extracted. First we determine the turn-on threshold voltage $V_{to}$ by measuring the current through the FET $I_{SD}$ as a function of the AG voltage $V_a$, while both BGs are kept at a constant voltage $V_b = 2.7$~V. This $V_b$ value set above the highest expected pinch-off voltage $V_{po}$ of the grid to avoid constricting the channel but low enough to minimize accumulation under the BGs. Then $V_{to}$ is automatically extracted by finding the voltage at which the curve overcomes the current noise threshold of 100 pA. Similarly, we determine the pinch-off voltages $V_{po}$ by measuring $I_{SD}$ as a function of $V_{b,left}$ and $V_{b,right}$ independently, while the other BG is kept at $2.7$~V. The AG is kept at 1~V above the previously determined $V_{to}$ to ensure a fully formed channel, which enables comparing the $V_{po}$ of FETs with differing $V_{to}$. The values for $V_{po}$ can again be estimated automatically through current thresholds, which are set at 20\% of the maximum current to account for FETs where the current does not go to 0 immediately after pinch-off, likely due to an undesired current path under the other barrier. FETs with pinch-off curves that are not sharp are not considered in further analysis. As expected based on the FET design, $V_{po} < V_{to}$ and the pinch-off curves are sharper than the turn-on curves due to the smaller distance from the BG to the 2DEG and the smaller area of 2DEG under the BG.

As a proof of principle, we show that the multi-gate FET supports a SET by lowering BG voltages to pinch off the accumulated channel and form tunnel barriers. Coulomb diamonds emerge in bias spectroscopy (Fig. \ref{fig:measurement}b) and from the height and width of a Coulomb diamond we extract a typical charging energy of $E_C = 8$ meV and a lever arm of $\alpha = 0.2$ eV/V. These metrics are valuable for characterizing and optimizing the material's suitability for spin qubit fabrication. Here we focus, as examples of possible routine characterization, on statistical measurements of the crossbar that can be incorporated in a fast fabrication-measurement cycle.

By repeating turn-on and pinch-off measurements as in Fig.~\ref{fig:measurement}a across all FETs of the crossbar, we achieve color-scale maps of $V_{to}$ and $V_{po}$, visualized in Fig.~\ref{fig:measurement}c-d according to physical location of selected unit cell in the crossbar. The crossbar achieved 100\% yield, meaning all 648 FETs were turned-on and pinched-off at $T = 1.7$~K. This large number of FET turn-ons and pinch-offs enables statistical analysis to determine how these voltages are affected by the gate dimensions and quantify uniformity of the material and fabrication at multiple length scales. In Fig.~\ref{fig:measurement}c we identify a clear vertical gradient of increasing turn-on voltages towards the bottom of the grid caused by the geometrical variations in the unit cell designs. 

A notable feature of the pinch-off plot in Fig.~\ref{fig:measurement}d are the vertical lines of similarly low pinch-off voltages located at odd BGs (11, 23, 27), corresponding to right barrier gates of the FETs. We speculate this effect arises from fabrication imperfections of the barriers. In addition, the operation of the device also contributes to asymmetry in the measurements. Since the right barrier pinch-off is measured last, this measurement is most affected by any device instability such as hysteresis. Due to the interleaved ohmic design, the device is robust against local errors such as fabrication imperfections, gate discontinuities or shorts between gate layers that can prevent a 2DEG from forming. Instead of failing to measure large quadratic sections of the crossbar, the consequences are limited to a linear loss of measurable unit cells along the row or column and the remaining unit cells are unaffected. Also the varying FET dimensions have little effect on the pinch-off voltages except where $W < 50$ nm, located at the bottom 7 rows of the grid. Since the positional variance is larger than trends caused by design differences for most unit cells, device-scale uniformity over length scales up to 230 $\mu$m can be estimated.

\begin{figure}%
    \includegraphics[width=89mm]{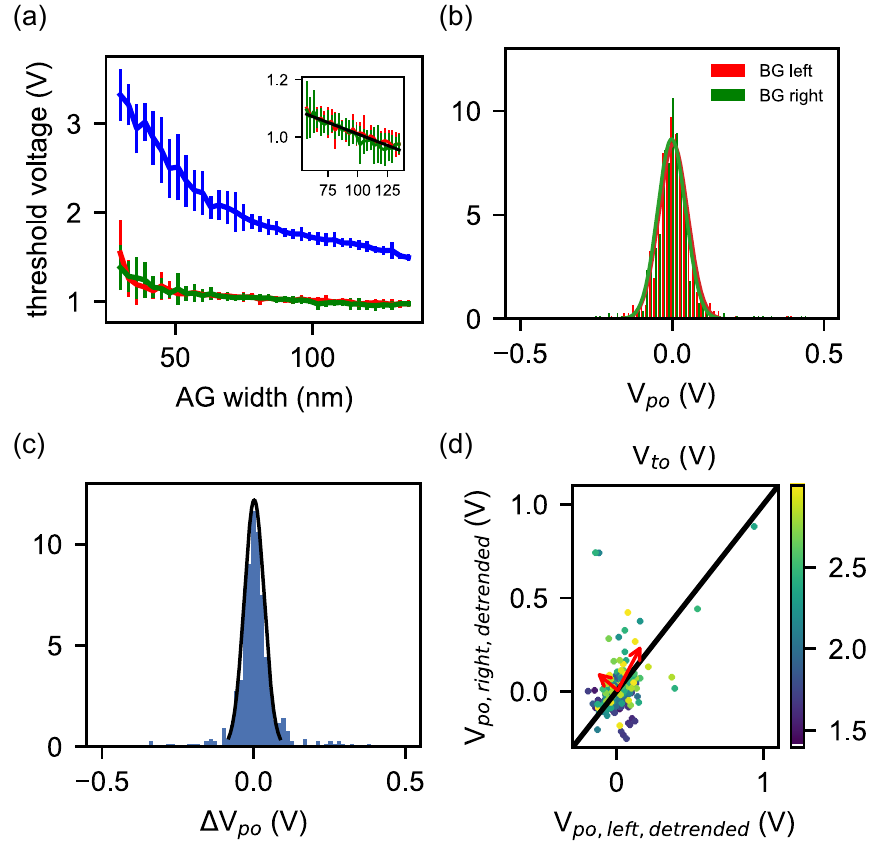}%
    \caption{(a) Turn-on and pinch-off voltages as a function of accumulation gate width. Each point is an average over a row of unit cells in Fig.~\ref{fig:measurement}c with identical AG width $W$ but differing BG spacing $L$. The inset is an enlargement of the pinch-off voltages where the associated turn-on voltage is always below 3V, with a linear fit (black line). (b) Probability density histogram of the pinch-off voltages measured throughout the grid where $V_{to} < 3$V, without the linear gate geometry trend. The colored lines are Gaussian fits for both distributions. (c) Probability density histogram of the pinch-off differences $\Delta V_{po}$ (blue) between the right and left barriers within the same FET and a Gaussian fit thereof (black line). (d) Pinch-off values of both barriers for each FET plotted against each other with the linear gate geometry trend removed. The principal components of the data are depicted as red arrows and a diagonal black line is a guide for the eye. }
\label{fig:analysis}
\end{figure}

The correlation between device design and behavior can be analyzed in our device, as the crossbar includes differences in gate geometries in every unit cell. While the barrier spacing was varied, no horizontal gradient can be seen in Fig.~\ref{fig:measurement}c-d, which indicates this parameter has no major influence on the turn-on or pinch-off. Evidently, the individual thresholds are primarily defined by the local gate with low voltage rather than the gate in full accumulation situated nearby. Therefore, the values of each row can be averaged to produce statistics on the effect of the accumulation gate width, as seen in Figure~\ref{fig:analysis}a. The increase in turn-on voltage with decreasing AG width is expected due to the narrow channel effect of MOSFETs\cite{Kroell1976}.

The pinch-off voltage behavior can be separated into two domains. Firstly, the increased pinch-off of FETs with $W < 50$ nm coincides with the appearance of turn-on values over 3V. In this case, the procedure of accumulating a proper channel at 1V above the turn-on threshold was not possible due to the setup output voltage limit of 4V. Therefore, the BG is partially acting as accumulation gate for the area around the crossing gates. Secondly, FETs with $W > 50$ nm exhibit a much weaker dependence, as seen in the inset of Fig.~\ref{fig:analysis}a. The large amount of measured FET enable obtaining a reliable linear fit to the data, despite constant behavior not being excluded by the standard deviation.

In Fig.~\ref{fig:analysis}b we visualize the pinch-off distributions across the grid as probability density histograms. In this analysis we consider only the FETs that turned-on with $V_{to} < 3$V and we take into account the linear trend related to the AG width observed in Fig.~\ref{fig:analysis}a by subtracting the linear fit, which includes the $1.17$~V average, from the data. The distributions of $V_{po}$, are well fitted with Gaussian distributions characterized by standard deviations of $47.5$~mV and $46.1$~mV for the left and right barriers, respectively. Since the FETs within the crossbar are positioned uniformly over an area of 200$\times$100~$\mu$m$^2$, standard deviation is a metric for quantifying uniformity of the disorder potential over the macroscopic scale. A relevant benchmark for this metric is the quantum dot charging energy, as proposals for scaling rely on specific levels of charge state uniformity. Comparing to the multi-electron occupancy SET in Fig.~\ref{fig:measurement}b, we find $\frac{\alpha \sigma_{V_{po}}}{E_C} = 1.17$ as the average standard deviation normalized to the charging energy. Considering the pinch-off gate representative of other gates in the crossbar, including the gate controlling the dot potential, we can estimate the variance of chemical potentials in SETs with the same applied voltage. The probability for the SET potential to be within the $E_C$ window required to be in the desired charge state associated with a certain voltage can be computed using the area under the Gaussian of the fit:

\begin{equation}
P_\text{desired state} = \int_{-\frac{E_C}{2 \alpha}}^{+\frac{E_C}{2 \alpha}} \frac{1}{\sigma_{V_{po}} \sqrt{2 \pi}} e^{\frac{-1}{2}\frac{x^2}{ \sigma_{V_{po}}^2}} dx,
\end{equation}
which equals 33.1\%.

Due to the symmetry of the barriers in each FET, the uniformity at the nanoscale can also be studied in our crossbar. The difference between the pinch-offs in one unit cell $\Delta V_{po} = V_{po,right} - V_{po,left}$ is shown in Fig.~\ref{fig:analysis}c, again fitted with a Gaussian. The standard deviation is $32.7$~mV, which indicates that correlation at the nanoscale, characterized by the barrier separation $L$ length scale within a single FET, is significantly larger than correlation at $\mu$m-scale, characterized by the spatial separation of different FETs. Assuming both pinch-offs within the same FET are sampled from the same distribution, since their environment is similar, the equivalent single pinch-off standard deviation is $32.7 \text{ mV}/\sqrt{2} = 23.1$ mV. However, normalized to the charging energy, the standard deviation $\frac{\alpha \sigma_{V_{po}}}{E_C} = 0.58$ corresponds to 61.3\% SETs with the desired charge state, which statistically reinforces how critical improving the material and fabrication uniformity is for realizing scalable qubits featuring shared gates. Moving towards advanced industrial processing is expected to yield uniformity improvements\cite{Giles2015}, but not by orders of magnitude. The metric can be further improved by making available energy states of the quantum dots more difficult to be filled, for example by increasing the charging energy through more confinement. 

The relevant uniformity length scale can be further investigated by plotting the pinch-offs of each FET against each other as seen in Fig.~\ref{fig:analysis}d. In the data analysis we again take into account the linear trend. The remaining variability is predominantly related to uniformity, where the shape is determined by the length scale of the dominant disorder. Disorders with characteristic length scales smaller than the barrier spacing of 30 - 130 nm or larger than the minimum distance between FETs of 3 $\mu$m contribute a 2D Gaussian variance because either the measurements are all uncorrelated or all correlated. On the other hand, disorders with length scales within this range are expected to primarily contribute to the diagonal because only the barriers within a FET are expected to be correlated. Therefore, principal component analysis\cite{Pearson1901} can be applied to quantify the direction-dependent variance of the distribution by determining the eigenvectors of the covariance matrix, as depicted the red arrows in Fig.~\ref{fig:analysis}d. The variance along the diagonal is 2.8 times as large as in the orthogonal direction, indicating the length scale we are probing is indeed the same magnitude as the dominant disorder.

In summary, we demonstrated a 2D crossbar controlled by cryogenic CMOS electronics with sublinear scaling of interconnects. As a result, we measured 648 multi-gated FETs in a single cooldown. All measurements are completely independent, with each unit cell covering a unique current path through the 2DEG, allowing for direct comparisons. This architecture is a powerful platform for analyzing device designs and their effect on device behavior by fabricating many devices with incremental differences. In this work the turn-on voltage dependence and pinch-off voltage independence on accumulation gate width was determined. Furthermore, statistical data can be obtained on the material and fabrication stack to assess the uniformity at various length scales using metrics that are relevant for spin qubit devices. With the cooldown bottleneck mitigated, the throughput of device measurements is now limited by the measurements themselves. Beyond hardware optimization, we envision that large amounts of data offer opportunities for training machine learning algorithms to improve tuning overhead. Finally, similar architectures, applicable to other material stacks, can open the door to successful experiments featuring low-yield structures as unit cells by leveraging quantity to achieve quality.

\bibliographystyle{naturemag}

\bibliography{bibliography.bib}

\end{document}